\documentclass[onecolumn, 12pt, journal]{IEEEtran}
\usepackage{amsmath,amsfonts, amssymb, mathrsfs, amsthm}
\usepackage[dvips]{graphicx}
\usepackage[dvips]{color}
\usepackage{pst-node}
\usepackage{latexsym}

\theoremstyle{plain}
\newtheorem{theorem}{Theorem}[section]

\theoremstyle{definition}
\newtheorem{definition}[theorem]{Definition}
\newtheorem{remark}[theorem]{Remark}
\newtheorem{lemma}[theorem]{Lemma}
\def\BC{\text{BC}}
\def\MAC{\text{MAC}}
\def\C{\mathcal{C}}
\def\N{\mathcal{N}}
\def\L{\Lambda}
\def\SNR{\text{SNR}}

\def\sp{\text{ }}
\def\tr{\text{Tr}}
\def\wf{\text{wf}}
\def\ep{\text{ep}}
\newcommand{\ol}[1]{\overline{#1}}
\def\PP{\mathbf{P}}
\def\EE{\mathbf{E}}
\begin{document}

\title{Achieving the Capacity of the $N$-Relay Gaussian Diamond Network Within $\log N$ Bits}

\author{\IEEEauthorblockN{Bobbie Chern and Ayfer \"{O}zg\"{u}r }\\
}


\maketitle

\begin{abstract}
We consider the $N$-relay Gaussian diamond network where a source node
communicates to a destination node via $N$ parallel relays through a cascade of a Gaussian broadcast (BC) and a multiple access (MAC) channel. Introduced in 2000 by Schein and Gallager, the capacity of this relay network is unknown in general. The best currently available capacity approximation, independent of the coefficients and the SNR's of the constituent channels, is  within an additive gap of $1.3 N$ bits, which follows from the recent capacity approximations for general Gaussian relay networks with arbitrary topology. 

In this paper, we approximate the capacity of this network within $2\log N$ bits. We show that two strategies can be used to achieve the information-theoretic cutset upper bound on the capacity of the network up to an additive gap of $O(\log N)$ bits, independent of the channel configurations and the SNR's. The first of these strategies is simple partial decode-and-forward. Here, the source node uses a superposition codebook to broadcast independent messages to the relays at appropriately chosen rates; each relay decodes its intended message and then forwards it to the destination over the MAC channel. A similar performance can be also achieved with  compress-and-forward type strategies (such as quantize-map-and-forward and noisy network coding) that provide  the $1.3 N$-bit approximation for general Gaussian networks, but only if the relays  quantize their observed signals at a resolution inversely proportional to the number of relay nodes $N$. This suggest that the rule-of-thumb  to quantize the received signals at the noise level in the current literature can be highly suboptimal. 
\let\thefootnote\relax\footnotetext{The authors are with the Electrical Engineering Department of Stanford University, Stanford, California 94305, USA, emails: \{bgchern, aozgur\}@stanford.edu. The work of Bobbie Chern was supported by
the Department of Defense (DoD) through the National Defense Science \&
Engineering Graduate Fellowship (NDSEG) Program. The work of Ayfer \"{O}zg\"{u}r was supported through NSF CAREER award 1254786. This work was presented in part in the Information Theory Workshop in Lausanne in 2012 \cite{bobbie}.}
\end{abstract}

\section{Introduction}
Consider a Gaussian relay network where a source node communicates to a
destination  with the help of intermediate relay nodes. Characterizing the capacity of this network is a long-standing open problem in network information theory. The seminal work of Cover and El-Gamal \cite{EGC79} has established several basic achievability schemes for the single relay channel, such as decode-and-forward and compress-and-forward. 
Recently, significant progress has been made by generalizing the compress-and-forward strategy  to achieve the capacity of any Gaussian relay network within an additive gap that depends on the network only through the total number of relay nodes $N$ (or the total number of transmit and receive antennas when nodes are equipped with multiple antennas) \cite{wnif}, \cite{abbas},  \cite{AyferISIT10}, \cite{flows}, \cite{KramerITW11}. 
The fact that the gap to capacity  is independent of the channel gains, the SNR's and the exact topology of the network suggests that  compress-and-forward can be universally good for relaying across different channel configurations, SNR regimes and topologies. However, the dependence of the gap to $N$ limits the applicability of these results to small networks with few relays. The best currently available capacity approximation in \cite{abbas} is within $1.3N$ bits (per second per Hz) of the information-theoretic cutset upper bound on the capacity of the network. For typical spectral efficiencies, this gap can quickly exceed the cutset upper bound with increasing $N$. This raises the following question: can we develop relaying strategies with provably smaller gap to capacity, in particular smaller than the order of $N$?

To the best of our knowledge, currently there are no nontrivial examples of Gaussian $N$-relay networks for which the gap to capacity has been demonstrated to be smaller than linear in $N$, independent of the channel coefficients and the SNR. A trivial example one can think of is the general class of $N$-relay networks comprised of orthogonal point-to-point AWGN channels.  In this case, routing information over different paths combined with decode-and-forward at the relays trivially achieves the exact capacity of the network (see \cite{KoetterEffrosMedard} for a generalization of this fact to other traffic scenarios).\footnote{A similar question is raised and a better than linear in $N$ capacity approximation is provided in \cite{bobak} for a class of layered networks with ergodic i.i.d. fading coefficients by using an ergodic lattice alignment strategy. However here, we are interested in networks with arbitrary fixed channel coefficients.} However, this setup discards the two main challenges in wireless, broadcast and superposition of signals. 

\begin{figure}[!t]
\centering
\psset{unit=0.015in}
\begin{pspicture}(0,-5)(120,100)
\psset{linewidth=0.5mm}
\definecolor{darkgreen}{rgb}{.15,.35,.15}
\begin{small}

\rput(-5,40){\circlenode[fillstyle=solid,fillcolor=orange]{S}{{\small $S$}}}
\rput(60,100){\circlenode[fillstyle=solid,fillcolor=orange]{A1}{$1$}}
\rput(60,60){\circlenode[fillstyle=solid,fillcolor=orange]{A2}{$2$}}
\rput(60,30){\Large $\vdots$}
\rput(60,-5){\circlenode[fillstyle=solid,fillcolor=orange]{AN}{$N$}}
\rput(125,40){\circlenode[fillstyle=solid,fillcolor=orange]{D}{$D$}}

\ncline[linewidth=0.5mm, linestyle=solid]{->}{S}{A1} \Aput{\large ${h_{1s}}$}
\ncline[linewidth=0.5mm, linestyle=solid]{->}{S}{A2} \Bput{\large $h_{2s}$}
\ncline[linewidth=0.5mm, linestyle=solid]{->}{S}{AN} \Bput{\large $h_{Ns}$}

\ncline[linewidth=0.5mm, linestyle=solid]{->}{A1}{D} \Aput{\large $h_{1d}$}
\ncline[linewidth=0.5mm, linestyle=solid]{->}{A2}{D} \Bput{\large $h_{2d}$}
\ncline[linewidth=0.5mm, linestyle=solid]{->}{AN}{D} \Bput{\large $h_{Nd}$}

\end{small}
\end{pspicture}
\caption{The Gaussian $N$-relay diamond network.}
\label{diamondfig}
\end{figure}
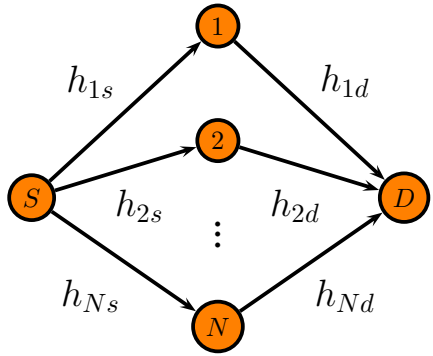

In this paper, we focus on the simplest setting of an $N$-relay Gaussian network that includes both broadcast and superposition, the $N$-relay diamond network. In this two-stage network, the source node is connected to $N$ relays
through a broadcast channel and the relays are connected to the
destination through a multiple-access channel. See Figure~\ref{diamondfig}. All received signals are corrupted by independent Gaussian noise.  The best currently available capacity approximation for this network, independent of the channel coefficients and the SNR, is  within an additive gap of $1.3 N$ bits, which follows from the capacity approximation for general Gaussian relay networks. 

In this paper, we provide $O(\log N)$-bit capacity approximations for this network. We first show that a simple modification of the compress-and-forward strategies (we take noisy network coding from \cite{abbas} as a reference) can reduce the gap to the information-theoretic cutset upper bound from $1.3N$  to $\log(N+1)+\log N+1$ bits. In the modified strategy, the relays quantize their received signals at a resolution inversely proportional to $N$. Equivalently, we let the power of the quantization noise introduced at each relay to increase linearly in $N$; the more relays we have, the more coarsely they quantize. The rule-of-thumb in the current literature is to quantize received signals at the noise level (independent of $N$), so that the injected quantization noise is more or less insignificant as compared to the Gaussian noise already corrupting the signals \cite{wnif}, \cite{abbas},  \cite{AyferISIT10}, \cite{flows}. However, this leads to a linear gap to the cutset upper bound. Our result reveals that there is a rate penalty for describing the quantized observations in compress-and-forward, and this penalty can be significantly larger than the rate penalty associated with coarser quantization. Follow-up work \cite{ritesh} has shown that this insight can be used to obtain tighter approximations for a much larger class of Gaussian relay networks.

We next show that  a similar performance can be obtained by a partial decode-and-forward strategy. Here, the source uses superposition coding to transmit independent messages to each of the relays at appropriately chosen rates; relays decode their intended messages, re-encode and forward them to the destination over the
multiple-access channel. A priori, one could expect this strategy to rather yield a linear  rate gap in $N$  to the cutset upper bound. Using a superposition codebook induces a rate penalty with respect to an i.i.d. Gaussian codebook since each message is decoded by treating some of the other messages as additional noise. Since for certain values of the channel coefficients in the broadcast phase, we may need to use an $N$-level superposition codebook (and since the undecoded messages constitute additional noise for the desired message at each relay except for the strongest relay which can decode all the messages), one may expect a constant rate loss associated with each message giving rise to a linear total rate loss with respect to the cooperative upper bound. Perhaps surprisingly, we  show that for all channel configurations and SNR's we can always find a rate point in the intersection of the broadcast and multiple access capacity regions such that the sum rate of the messages is only $2\log N$ bits away from the information-theoretic cutset upper bound. The key ingredient we use is the Edmond's polymatroid intersection theorem.

The rest of the paper is organized as follows. In Section~\ref{sec:model}, we formally introduce the  model for the diamond relay network. Section~\ref{sec:mainres} provides a summary of our main results. Sections \ref{sec:cf} and \ref{sec:pdf} include the proofs of our main results by concentrating on the compress-and-forward and partial decode-and-forward strategies respectively. Section~\ref{sec:sim} provides a discussion of our conclusions based on numerical evaluations. The appendix contains an extension of our results to the case with multiple antennas.
  
\subsection{Related Work on the Diamond Network}
 
The Gaussian diamond relay network was introduced by Schein and Gallager in \cite{schein, schein2}.  For the case when $N=2$, rates achievable by decode-and-forward and amplify-and-forward were analyzed in \cite{schein2}. In the asymptotic regime when $N\to\infty$, amplify-and-forward was  shown to be asymptotically optimal in \cite{gastpar}. The rate achieved by amplify-and-forward over the $N$-relay diamond network was also investigated in \cite{urssuhas} for the specific case when all channel coefficients are equal to each other and a constant additive approximation to the capacity of this symmetric setup was derived. \cite{kochman, rezaei} provided achievable schemes for Gaussian diamond networks with bandwidth mismatch, while \cite{xue, bagheri, sid} considered the diamond setting with half-duplex relays. \cite{simplification} provided a hybrid approximation for the capacity of the $N$-relay diamond network with smaller additive gap at the expense of also incurring a multiplicative gap to capacity. This hybrid approximation was based on using only a carefully chosen subset of the available relays.

It is now well-understood that while decode-and-forward and amplify-and-forward, the two most commonly considered strategies for the diamond network, can perform extremely well for specific channel configurations (for example,  amplify-and-forward with equal channel gains  \cite{urssuhas}), they can perform arbitrarily
away from capacity for other channel configurations. For example, \cite{simplification} shows that the best rate that can be achieved with amplify-and-forward in any $N$-relay diamond network is approximately equal to the rate achieved by using only the best relay, which
can in turn be as small as half the capacity of the whole network. Therefore,
amplify-and-forward cannot provide a constant gap approximation to the capacity
across different channel parameters and SNR's, such as the $O(\log N)$
approximation provided by the strategies in this paper. Prior to this work, the best uniform capacity approximation for the diamond network, over all channel coefficients and SNR's,  was the $1.3N$-bit additive approximation provided in \cite{abbas} for general Gaussian networks.\footnote{ A version of the
$O(\log(N))$gap with quantize-map-and-forward was also presented in \cite{ayan}
independently at the same conference as our work \cite{bobbie}.} 

\section{Model}\label{sec:model}
We consider the Gaussian $N$-relay diamond network depicted in Fig.
\ref{diamondfig}, where the source node $s$ wants to communicate to the
destination node $d$ with the help of $N$ relay nodes, denoted
$\mathcal{N}=\{1,\dots, N\}$.    Let $X_s[t]$ and $X_i[t]$ denote the signals transmitted
by the source node and the relay node $i \in \mathcal{N}$ respectively at time
instant $t \in \mathbb{N}$.  Similarly, $Y_d[t]$ and $Y_i[t]$ denote the
signals received by the destination node and the relay node $i$ respectively.
These signals are related as 
\begin{align}
Y_i[t] = h_{is}X_s[t] + Z_i[t], \label{eq:BCmod}\\
Y_d[t] = \sum_{i=1}^N h_{id}X_i[t] + Z[t],\label{eq:MACmod}
\end{align}
where $h_{is}$ denotes the complex channel coefficient between the source and
relay node $i$, and $h_{id}$ denotes the complex channel coefficient between
the relay node $i$ and the destination node. We assume the fixed channel coefficients are known to all the nodes in the network. $Z_i[t]$ and $Z[t]$ are independent
and identically distributed circularly symmetric Gaussian random variables of
variance $\sigma^2$. All transmitted signals are subject to an average power constraint $P$ and  we define $$\SNR=P/\sigma^2.$$  Note that the equal power constraint
assumption is without loss of generality as the channel coefficients are
arbitrary.

The capacity of this network is defined as the largest rate at which $s$ can reliably communicate to $d$ in the following standard way: Let $W$ denote the message $s$ wants to communicate to $d$. Assume $W$ is uniformly distributed over $\{1,\dots, \lceil 2^{TR}\rceil\}$ for some integer $T$ and $R\geq 0$. A blocklength $T$ and rate $R$ code is a collection of functions   $f:\{1,\dots, \lceil 2^{TR}\rceil\}\to \mathbb{C}^T$, $f_i:\mathbb{C}^T\to \mathbb{C}^T$ for $i=1,\dots, N$, and $g:\mathbb{C}^T\to \{1,\dots, M\}$. The  encoding  function $f$ maps the message at the source to a block of $T$ channel inputs, 
\begin{equation}\label{eq:cap1}
(X_s[1],\dots, X_s[T])=f(W).
\end{equation} The mapping function $f_i$ maps a block of $T$ channel outputs at relay $i$ to a block of $T$ channel inputs,
 \begin{equation}\label{eq:cap2}(X_i[1],\dots, X_i[T])=f_i(Y_i[1],\dots, Y_i[T]).\end{equation} The decoding function $g:\mathbb{C}^T\to \{1,\dots, M\}$  maps a block of channel observations at the destination to a guess $\hat{W}$ for the transmitted message, \begin{equation}\label{eq:cap3}\hat{W}=g(Y_d[1],\dots, Y_d[T]).\end{equation} The code satisfies an average power constraint $P$ if $$ \sum_{t=1}^T \EE\left[|X_s[t]|^2\right]\leq P\quad \text{and}\quad \sum_{t=1}^T \EE\left[|X_i[t]|^2\right]\leq P,$$ for $i=1,\dots, N$ and has an average probability of error $\PP[\hat{W}\neq W]$. 
 
A rate $R$ is said to be achievable if  there exists
a sequence of codes of blocklength $T$ and rate $R$ that satisfy an average power constraint $P$ and the average probability of error $\PP[\hat{W}\neq W]\to 0$ as $T\to\infty$. The capacity $C$ of the diamond network is the largest achievable rate $R$.

When we consider the case when nodes are equipped with multiple antennas in the appendix, we will prefer to denote the channel matrices with capital letters. In this case, we will assume that the
source node $s$ has $n_s \ge 1$ transmit antennas, the destination node $d$ has
$n_d \ge 1$ receive antennas, and relay $i$ has $n_i \ge 1$ transmit and
receive antennas. The relation between the channel inputs and outputs is denoted by
\begin{align*}
Y_i[t] &= H_{is}X_s[t] + Z_i[t], \quad i\in \mathcal{N}\\
Y_d[t] &= \sum_{i=1}^N H_{id}X_i[t] + Z[t],
\end{align*}
in this case, where $H_{is} \in \mathbb{C}^{n_i \times n_s}$ is the channel matrix between
the source node $s$ and relay $i$, $H_{id} \in \mathbb{C}^{n_d \times n_i}$ is
the channel matrix between relay $i$ and the destination node $d$. Note that in this case the channel input and output signals are complex vectors of appropriate dimension and $Z_i[t]$ and $Z(t)$ are circularly symmetric Gaussian random vectors with covariance $\sigma^2I$ where $I$ is the identity matrix of appropriate dimension. We still assume an equal power constraint
$P$ at the notes, which in this case amounts to
$$
\lim_{T\to\infty} \frac{1}{T}\sum_{t=1}^T \EE\left[||X_i[t]||^2\right]\leq P.
 $$
The capacity of this network is defined analogously to the scalar case. To simplify the statement of our results we assume the number of antennas at the source and the number of antennas at the destination are smaller than the total number of antennas at the relays, i.e. $n_s\leq \sum_i n_i$ and $n_d\leq \sum_i n_i$, however the analysis also holds for the general case. 

\bigbreak

Although not directly part of our problem, in the sequel we will be interested in the capacity regions of the broadcast channel (BC) from the source node to the relays and the multiple-access channel (MAC) from the relays to the destination. We next define these two channels:

\begin{figure}[!t]
\begin{center}
\psset{unit=0.015in}
\begin{pspicture}(-5,-5)(120,100)
\psset{linewidth=0.5mm}
\definecolor{darkgreen}{rgb}{.15,.35,.15}
\begin{small}

\rput(-65,40){\circlenode[fillstyle=solid,fillcolor=orange]{S}{{\small $S$}}}
\rput(0,100){\circlenode[fillstyle=solid,fillcolor=orange]{A1}{$1$}}
\rput(0,60){\circlenode[fillstyle=solid,fillcolor=orange]{A2}{$2$}}
\rput(0,30){\Large $\vdots$}
\rput(0,-5){\circlenode[fillstyle=solid,fillcolor=orange]{AN}{$N$}}

\rput(120,100){\circlenode[fillstyle=solid,fillcolor=orange]{A12}{$1$}}
\rput(120,60){\circlenode[fillstyle=solid,fillcolor=orange]{A22}{$2$}}
\rput(120,30){\Large $\vdots$}
\rput(120,-5){\circlenode[fillstyle=solid,fillcolor=orange]{AN2}{$N$}}
\rput(185,40){\circlenode[fillstyle=solid,fillcolor=orange]{D}{$D$}}

\ncline[linewidth=0.5mm, linestyle=solid]{->}{S}{A1} \Aput{\large ${h_{1s}}$}
\ncline[linewidth=0.5mm, linestyle=solid]{->}{S}{A2} \Bput{\large $h_{2s}$}
\ncline[linewidth=0.5mm, linestyle=solid]{->}{S}{AN} \Bput{\large $h_{Ns}$}

\ncline[linewidth=0.5mm, linestyle=solid]{->}{A12}{D} \Aput{\large $h_{1d}$}
\ncline[linewidth=0.5mm, linestyle=solid]{->}{A22}{D} \Bput{\large $h_{2d}$}
\ncline[linewidth=0.5mm, linestyle=solid]{->}{AN2}{D} \Bput{\large $h_{Nd}$}

\end{small}
\end{pspicture}
\caption{A broadcast and a multiple access channel.}
\label{BCandMAC}
\end{center}
\end{figure}
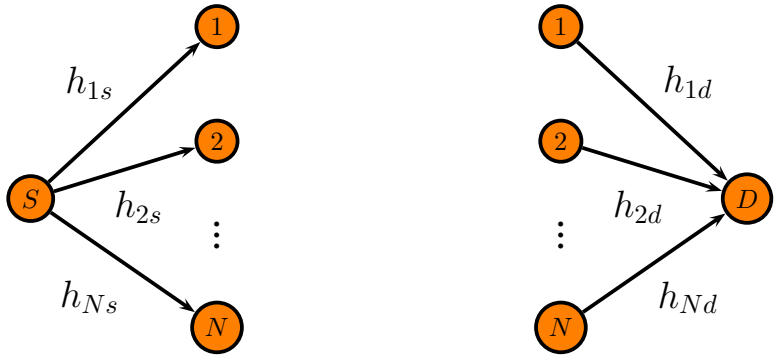

\subsection{BC Channel}\label{sec:BCmod} Consider a communication system where a sender $s$ has  $N$ independent messages $W_1,\dots, W_N$ to communicate to $N$ destinations as depicted in the left figure in Fig.~\ref{BCandMAC}. Each destination $i$ is only interested in its corresponding message $W_i$. This is called a broadcast channel. A code of blocklength $T$ and rate $(R_1, \dots, R_N)$ for communicating the $N$ messages $W_1,\dots, W_N$, where $W_i$ is uniformly distributed over $\{1,\dots, \lceil 2^{TR_i}\rceil\}$, to their respective destinations is defined analogously to \eqref{eq:cap1}, \eqref{eq:cap2}, \eqref{eq:cap3} as a set containing an encoding function at the source (satisfying the power constraint $P$) and $N$ decoding functions, one for each destination. The capacity region $\mathcal{C}_{BC}$ is the closure of the set of achievable rates $(R_1, \dots, R_N)$. See Chapter~5 of \cite{abbasbook} for formal definitions. 

In the sequel, we will be interested in the broadcast channel induced by the first stage of the diamond network in Fig.~\ref{BCandMAC}. Here, the relays  act as destinations for $N$ independent messages from the source and  the channel input and outputs are related by \eqref{eq:BCmod}. We denote the capacity region of this channel by  $\mathcal{C}_{BC}^{s\to\mathcal{N}}$. In this Gaussian case, the capacity region $\mathcal{C}_{BC}^{s\to\mathcal{N}}$ is exactly characterized (see \cite[Theorem 5.3]{abbasbook}).

\subsection{MAC Channel}\label{sec:MACmod} Consider a communication system where $N$ senders want to simultaneously communicate to a destination as depicted in the right figure in Fig.~\ref{diamondfig}. Each sender has an independent message $W_i$ to communicate to the destination node. This is called a multiple-access channel. A code of blocklength $T$ and rate $(R_1, \dots, R_N)$ for communicating the $N$ messages $W_1,\dots, W_N$ of the senders, where $W_i$ is uniformly distributed over $\{1,\dots, \lceil 2^{TR_i}\rceil\}$, is defined analogously to \eqref{eq:cap1}, \eqref{eq:cap2}, \eqref{eq:cap3} as a set of encoding functions at the senders and a decoding function at the destination. The capacity region $\mathcal{C}_{MAC}$ is the closure of the set of achievable rates $(R_1, \dots, R_N)$. See Chapter~4 of \cite{abbasbook} for formal definitions. The capacity region of the MAC channel has been completely characterized (see  in  \cite[Theorem 4.4]{abbasbook}).

In the sequel, we will be interested in two MAC channels induced by the two stages of the diamond network in Fig.~\ref{BCandMAC}. The capacity regions of these two MAC channels will be denoted by $\mathcal{C}_{MAC}^{\mathcal{N}\to d}$ and  $\mathcal{C}_{MAC}^{\mathcal{N}\to s}$. In the first case,  we will assume that each relay has an independent message to communicate to the destination and  the channel input and outputs are related by \eqref{eq:MACmod}. Here, each relay is subject to a power constraint $P$. In the second case, we will assume that each relay has an independent message to communicate to the source node and  the channel input and output relations are given by the inverse channel of \eqref{eq:BCmod}, i.e.,
\begin{equation}\label{eq:inverseBC}
Y_s[t] = \sum_{i=1}^N h_{is}X_i[t] + Z_s[t]
\end{equation} 
where $Z_s[t]$ is circularly-symmetric Gaussian noise with variance $\sigma^2$. When the relays are subject to average power constraints $P_1,\dots, P_N$ respectively, we will denote the corresponding capacity region by $\mathcal{C}_{MAC}^{\mathcal{N}\to s}(P_1,\dots, P_N)$.

\section{Main Result}\label{sec:mainres}
The main conclusions of this paper are summarized in the following theorems.

\begin{theorem}\label{thm:nnc}Let $\ol{C}$ be the information-theoretic cutset upper bound on the capacity of the $N$-relay diamond network. Noisy network coding at the relays can achieve a rate 
\begin{equation}
R_{NNC} \ge \ol{C} - G_1,
\end{equation}
where $G_1=\log(N+1)+\log N+1$  when nodes have single antennas.
\end{theorem}
\begin{remark}\label{rem:nnc} When nodes have multiple antennas the gap becomes
$$G_1=n_s \log \left( M + 1\right) + n_s \log (1 +
\frac{n_s -1}{n_a}) + n_d \log (\max_{i \in \N} n_i) + n_d \log (1 +
\frac{M - 1}{n_b})+1,$$ where $M=\sum_{i \in \N} n_i$, $n_a = \min(n_s, n_1, \ldots, n_N)$,
$n_b = \min(n_d, n_1, \ldots n_N)$. Note that the gap increases linearly in the number of antennas at the source and the destination and logarithmically in the total number of antennas at the relays. When all nodes have a single antenna, the gap reduces to $\log(N+1)+\log N+1$.
\end{remark}

\begin{theorem}\label{thm:pdf}
A partial decode-and-forward strategy at the relays achieves a rate
\begin{equation}
\label{eq:pdf_gap}
R_{PDF} \ge \ol{C} - G_2,
\end{equation}
where $G_2= 2\log N$ in the case of single antenna nodes. 
\end{theorem}

\begin{remark}\label{rem:pdf}  When nodes have multiple antennas the gap becomes
$$G_2 = n_s \log \left( M \right) + n_s \log (1 +
\frac{n_s -1}{n_a}) + n_d \log \left( \max_{i \in \N} n_i \right) + n_d \log (1
+ \frac{M - 1}{n_b})$$ where $M$, $n_a$ and
$n_b$ are defined as before.  Note that when all nodes have a single antenna, the gap reduces to $2 \log N$.
\end{remark}

We prove the two theorems in the following two sections. The extensions to multiple antennas in the two remarks are given in the appendix.

\section{Noisy Network Coding}\label{sec:cf}
In this section, we prove Theorem~\ref{thm:nnc} by investigating the performance of compress-and-forward based strategies for the diamond network. We take the noisy network
coding result in \cite{abbas} as a reference, however the discussion applies to other compress-and-forward based strategies such as the quantize-map-and-forward in \cite{wnif}, which was the first strategy to provide $O(N)$-bit approximations for the capacity of Gaussian networks. The main idea of these strategies is that relays quantize their received signals without decoding and independently map them to Gaussian codebooks. It has been more recently shown that a similar performance can be also achieved with classical compress-and-forward, where the quantized signals at the relays are binned before transmission at appropriately chosen rates, and they are decoded
successively before decoding the actual source message \cite{flows,Xie,KramerITW11}.

The performance achieved by noisy network coding is given in \cite[Theorem 1]{abbas} as 
\begin{align}\label{ratennc}
R_{NNC}=\min_{\L\subseteq \mathcal{N}} I(X_s,X_\L; Y_d, \hat{Y}_{\bar{\L}}|X_{\bar{\L}}) - I(Y_\L; \hat{Y}_\L | X, X_\mathcal{N}, \hat{Y}_{\bar{\L}}, Y_d).  
\end{align}
for some joint probability distribution $\prod_{i\in\mathcal{N}}
p(x_i)p(\hat{y_i}|y_i, x_i)$ where $X_\L=\{X_i, i\in\L\}$, $\ol{\L}=\mathcal{N}\setminus \L$ and $Y_{\ol{\L}}, X_{\ol{\L}}$ are defined analogously.

Comparing this with the information-theoretic cutset upper bound on the capacity of the network given by
\cite{eit}
\begin{equation}\label{eq:cutset}
\ol{C}=\sup_{X_s,X_1,\dots, X_N} \min_{\L\subseteq \mathcal{N}} I(X_s, X_\L; Y_d, Y_{\ol{\L}}\,|\, X_{\ol{\L}}),
\end{equation}
we observe the following differences. The first term in \eqref{ratennc} is similar to \eqref{eq:cutset} but with $Y_{\ol{\L}}$ in \eqref{eq:cutset} replaced by  $\hat{Y}_{\bar{\L}}$ in \eqref{ratennc}. The difference corresponds to a rate loss due to the quantization noise introduced by the relays. Second, while the maximization in \eqref{eq:cutset} is over all possible input distributions, only independent input distributions are  admissible  in \eqref{ratennc}. This corresponds to rate loss with respect to a potential beamforming gain accounted for in the upper bound. Third, there is the extra term $I(Y_\L; \hat{Y}_\L | X, X_\mathcal{N}, \hat{Y}_{\bar{\L}}, Y_d)$ reducing the rate in \eqref{ratennc}. This corresponds to the rate penalty for communicating the quantized observations to the destination along with the desired message.

The works in the current literature \cite{wnif, abbas, AyferISIT10} choose $X_i$ in \eqref{eq:cutset} to be i.i.d. circularly symmetric Gaussian of variance $P$ and 
$$
\hat{Y}_i=Y_i+\hat{Z}_i, \qquad i\in\mathcal{N},
$$
where $\hat{Z}_i,\,i\in\mathcal{N}$ are i.i.d. circularly symmetric and complex
Gaussian random variables of variance $\sigma^2$ independent of everything else. This results in $O(\log N)$ difference between the first term of \eqref{ratennc} and \eqref{eq:cutset} while the second term in 
\eqref{eq:cutset} is  $O(N)$, resulting in an overall gap of $O(N)$. 

To reduce the $O(N)$ rate loss for communicating the quantized observations, we can instead quantize at a coarser resolution, i.e. take the variance of $\hat{Z}_i$ to be $N\sigma^2$. Then, the first mutual information becomes
\begin{align*}
I(X_s, X_\L;\hat{Y}_{\ol{\L}}, Y_d | X_{\ol{\L}}) &= I(X_s, X_\L;\hat{Y}_{\ol{\L}} | X_{\ol{\L}}) + I(X_s,X_\L; Y_d | \hat{Y}_{\ol{\L}}, X_{\ol{\L}}) \\
&= I(X_s; \hat{Y}_{\ol{\L}} | X_{\ol{\L}}) + I(X_\L; \hat{Y}_{\ol{\L}} | X_{\ol{\L}}, X_s) + I(X_\L; Y_d | \hat{Y}_{\ol{\L}}, X_{\ol{\L}}) + I(X_s; Y_d | \hat{Y}_{\ol{\L}}, X_{\ol{\L}}, X_\L) \\
&\stackrel{(a)}{=}I(X_s; \hat{Y}_{\ol{\L}}) + I(X_\L; \sum_{i \in \L} h_{id}X_i + Z) \\
&\stackrel{(b)}{=} \log\big(1 + \sum_{i\in \ol{\L}}|h_{is}|^2\text{ SNR}/(N+1)\big) + \log\big(1 + \sum_{i\in \L}|h_{id}|^2\text{ SNR}\big),
\end{align*}
where (a) follows from the independence of the $X_i$'s  and the structure of the network and (b) follows by evaluating the mutual informations for the chosen distributions.  The second term in \eqref{ratennc} is now given by
$$
I(Y_\L; \hat{Y}_\L | X, X_\mathcal{N}, \hat{Y}_{\bar{\L}}, Y_d)=|\L|\log(1+\frac{1}{N})\leq \frac{|\L|}{N}\leq 1.
$$

We next bound the gap between the resultant rate and the cutset upper bound by first deriving a simple upper bound on the cutset bound. We have  
\begin{align}
\ol{C}&=\sup_{X_s,X_1,\dots, X_N} \min_{\L\subseteq \mathcal{N}} I(X_s, X_\L; Y_d, Y_{\ol{\L}}\,|\, X_{\ol{\L}}) \nonumber \\
&\stackrel{(a)}{\leq} \min_{\L\subseteq \mathcal{N}} \sup_{X_s,X_1,\dots, X_N} I(X_s, X_\L; Y_d, Y_{\ol{\L}}\,|\, X_{\ol{\L}}) \nonumber \\
&\stackrel{(b)}{\leq}\min_{\Lambda\subseteq [N]} \sup_{X_s,X_{\Lambda}} I(X_s, X_\Lambda; \sum_{i\in\Lambda} h_{id} X_i+Z, Y_{\overline{\Lambda}})\hspace*{4cm}\nonumber
\end{align}
\begin{align}
&\stackrel{(c)}{=} \min_{\L \subseteq \N} \sup_{X_s} I(X_s; Y_{\ol{\L}}) + \sup_{X_\L} I(X_\L; \sum_{i \in \L} h_{id} X_i + Z) \nonumber \\
&\stackrel{(d)}{=} \min_{\L \subseteq \N} \Bigg(\log \Big( 1 + \SNR \sum_{i \in \ol{\L}} |h_{is}|^2 \Big) + \log \Big( 1 + \SNR \big( \sum_{i \in \L} |h_{id}| \big)^2 \Big) \Bigg).\label{eq:cutset3}
\end{align}
Here, (a) follows by exchanging the order of min and sup; (b) follows because 
\begin{align*}
&I(X_s, X_\Lambda; Y_d, Y_{\overline{\Lambda}}\,|\, X_{\overline{\Lambda}})=I(X_s, X_\Lambda; Y_d-\sum_{i\in\overline{\Lambda}} h_{id} X_i, Y_{\overline{\Lambda}}\,|\, X_{\overline{\Lambda}})\\
&\,=h(Y_d-\sum_{i\in\overline{\Lambda}} h_{id} X_i, Y_{\overline{\Lambda}}\,|\, X_{\overline{\Lambda}})-h(Y_d-\sum_{i\in\overline{\Lambda}} h_{id} X_i, Y_{\overline{\Lambda}}\,|\,X_s, X_\Lambda, X_{\overline{\Lambda}})\\
&\,=h(Y_d-\sum_{i\in\overline{\Lambda}} h_{id} X_i, Y_{\overline{\Lambda}}\,|\, X_{\overline{\Lambda}})-h(Z, Z_{\overline{\Lambda}})\\
&\leq h(Y_d-\sum_{i\in\overline{\Lambda}} h_{id} X_i, Y_{\overline{\Lambda}})-h(Z, Z_{\overline{\Lambda}})\\
&=I(X_s, X_\Lambda; \sum_{i\in\Lambda} h_{id} X_i+Z, Y_{\overline{\Lambda}}).
\end{align*}
Note that this last expression maximized over all random variables $X_s, X_{\Lambda}$ is the capacity of the point to point channel between $\{s,\Lambda\}$ and $\{\overline{\Lambda},d\}$. The capacity of this channel can be further upper bounded by the sum of the capacities of the SIMO channel between $s$ and $\{\overline{\Lambda}\}$ and the MISO channel between $\{\Lambda\}$ and $d$ which is the result stated in (c). Formally, (c) follows because
\begin{align*}
&I(X_s, X_\Lambda; \sum_{i\in\Lambda} h_{id} X_i+Z, Y_{\overline{\Lambda}})\\
&\,\,\leq h(\sum_{i\in\Lambda} h_{id} X_i+Z)+ h( Y_{\overline{\Lambda}})-h(Z)- h(Z_{\overline{\Lambda}})\\
&\,\,=I(X_s; Y_{\overline{\Lambda}})+I(X_\Lambda; \sum_{i\in\Lambda} h_{id} X_i+Z).
\end{align*}
The solutions to the maximization of these mutual informations over the input distributions are well-know  and yield the capacities of the corresponding SIMO and MISO channels \cite{wireless}. \eqref{eq:cutset3} is obtained by plugging in these capacities.

It can be easily verified that the total gap of (\ref{ratennc}) to the upper bound in (\ref{eq:cutset3}) is  bounded by $\log(N+1)+\log N+1$. This completes the proof of Theorem~\ref{thm:nnc}.

\section{Partial Decode and Forward}\label{sec:pdf}
We consider a partial decode-and-forward strategy where the first stage of the
communication is treated as a broadcast channel and the second stage is treated
as  a multiple access channel. The source splits its message $W$ of rate $R_{PDF}$ into $N$ messages $W_1, \dots, W_N$ of corresponding rates $R_i$, $i=1,\dots,N$ such that $R_{PDF}=\sum_i R_i$. Relay $i$ can decode its corresponding message $W_i$ if the rates $R_i$, $i=1,\dots,N$ lie in the capacity region of the broadcast channel from the source to the relays. We denote this region (formally defined in Section~\ref{sec:BCmod}) by $\mathcal{C}_{BC}^{s\to\mathcal{N}}$. Once each relay decodes its message, it can re-encode and forward it to the destination. The messages $W_1, \dots, W_N$ can be simultaneously communicated to the destination node if their rates $R_i$, $i=1,\dots,N$  also lie in the capacity region of the MAC channel from the relays to the destination. We denote this region (formally defined in Section~\ref{sec:MACmod}) by $\mathcal{C}_{MAC}^{\mathcal{N}\to d}$.  With this relaying strategy, we can achieve any rate given by
\begin{equation}\label{eq:RPDF}
R_{PDF}=\sum_{i\in\mathcal{N}} R_i\qquad\text{s.t.}\qquad \{R_1,\dots, R_N\}\in\mathcal{C}_{BC}^{s\to\mathcal{N}} \cap \mathcal{C}_{MAC}^{\mathcal{N}\to d}.
\end{equation}
Clearly, to maximize the rate achieved by this strategy, we need to find the rate point $\{R_1,\dots,R_N\}\in\mathcal{C}_{BC}^{s\to\mathcal{N}} \cap \mathcal{C}_{MAC}^{\mathcal{N}\to d}$ with largest sum-rate. Without explicitly identifying this maximal point, we will show that for any value of the channel coefficients and the SNR there exists a rate point $\{R_1,\dots,
R_N\}\in\mathcal{C}_{BC}^{s\to\mathcal{N}} \cap \mathcal{C}_{MAC}^{\mathcal{N}\to d}$ such that the difference
between $\sum_{i\in\mathcal{N}} R_i$ and the information-theoretic cutset upper bound on the capacity of the network, $\bar{C}$, is bounded. To prove this, we will make use of Edmond's polymatroid intersection theorem. 

The region $C_{MAC}^{\mathcal{N}\to d}$ is known to have a polymatroid structure \cite{tsehanly}.
The region $\mathcal{C}_{BC}^{s\to\mathcal{N}}$ however is not polymatroidal. Below, we define a
polymatroid, and use the duality between the BC and MAC capacity regions
\cite{bcmac} to find a polymatroidal lower bound on the BC capacity region.  We
then use Edmond's polymatroid intersection (\cite{polymatroid}, Corollary
46.1b) to find an intersection point in the two polymatroid regions with
largest sum rate.

\begin{definition}\label{def1}
Let $f: 2^{\mathcal{N}}\rightarrow \mathbb{R}^+$ be a set function. The polyhedron 
$$
P(f):=\{(x_1,\dots, x_N): \sum_{i\in S} x_i\leq f(S), \forall S\subseteq\mathcal{N}, x_i\geq 0,\, \forall i\}
$$
is a polymatroid if the set function $f$ satisfies
\begin{enumerate}
\item $f(\emptyset)=0$ (normalized).
\item $f(S)\leq f(T)$ if $S\subseteq T$ (non-decreasing).
\item $f(S)+f(T)\geq f(S\cup T)+f(S\cap T)$ (submodular).
\end{enumerate}
\end{definition}

The MAC capacity region $\mathcal{C}_{MAC}^{\mathcal{N}\to d}$ is given by
$$
P(f)=\{(R_1,\dots, R_N): \sum_{i\in S} R_i\leq f(S), S\subseteq \mathcal{N}, R_i\geq 0,\, \forall i\}
$$ 
where 
$$
f(S)=\log\left(1 + \sum_{i\in S}|h_{id}|^2\text{ SNR}\right).
$$
Since $f$ satisfies the conditions in Definition~\ref{def1}, $P(f)$ is a
polymatroid \cite{tsehanly}. By the duality established in \cite{bcmac}, the BC capacity region
is given by
$$
C_{BC}^{s\to\mathcal{N}}=\bigcup_{(P_1\dots, P_N):\sum_{P_i}=P} \mathcal{C}_{MAC}^{\mathcal{N}\to s}(P_1,\dots,P_N)
$$
where $\mathcal{C}_{MAC}^{\mathcal{N}\to s}(P_1,\dots,P_N)$ is the capacity region of a MAC
channel from the relays to the source node with relay $i$ constrained to an average power $P_i$. This region has been formally defined in Section~\ref{sec:MACmod}. Any choice for the powers $P_1,\dots,P_N$ such that $\sum_i{P_i}=P$ provides a lower bound on the BC capacity region. In particular, $\mathcal{C}_{MAC}^{\mathcal{N}\to s}(P/N,\dots,P/N)\subseteq
\mathcal{C}_{BC}^{s\to\mathcal{N}}$, or equivalently,
$$
P(g) \subseteq \mathcal{C}_{BC}^{s\to\mathcal{N}}
$$ 
where
$$
g(S)=\log\left(1 + \sum_{i\in S}|h_{is}|^2\frac{\text{ SNR}}{N}\right).
$$
Clearly, $P(g)$ is also a polymatroid. It then follows from Edmond's
polymatroid intersection (\cite{polymatroid}, Corollary 46.1b) that
\begin{align*}
\max\left\{\sum_{i} R_i : (R_1,\dots, R_N) \in P(f)\, \cap\, P(g)\right\} = \min_{\L \in \mathcal{N}} \left\{f(\L) + g(\ol{\L}) \right\}.
\end{align*}
Therefore, partial decode-and-forward can achieve a rate
\begin{align}
R_{PDF}&= \min_{\L \in \mathcal{N}}\quad f(\L) + g(\ol{\L}) \nonumber \\
&= \min_{\L \in \mathcal{N}} \Big(\log\big(1 + \sum_{i\in \ol{\L}}|h_{is}|^2\frac{\SNR}{N}\big) + \log\big(1 + \sum_{i\in \L}|h_{id}|^2\text{ SNR}\big)\Big)\label{ratepdf}
\end{align}

By comparing \eqref{ratepdf} and \eqref{eq:cutset3}, it can be easily verified that $$R_{PDF}\geq \ol{C}-2\log N.$$
This completes the proof of Theorem~\ref{thm:pdf}.

\subsection{Discussion}
\begin{figure}[!t]
\centering
\includegraphics[width=5in]{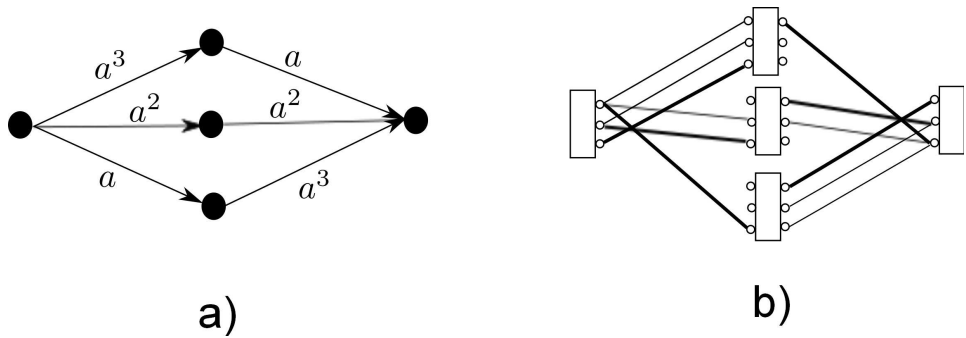}
\caption{A $3$-relay diamond network.}
\label{fig:fig2}
\end{figure}
The above argument proves the existence of a rate point $\{R_1,\dots,
R_N\}$ in the intersection of the BC and MAC capacity regions with sumrate within $2\log N$ bits of the cutset upper bound for any value of the channel coefficients. In this section, we aim to obtain more insight on the choice of the optimal rate point $\{R_1,\dots, R_N\}$ by concentrating on the example of a $3$-relay diamond network given in Fig.~\ref{fig:fig2} -(a). Here, the labels indicate the SNR's of the corresponding links (assume the transmit and noise powers are normalized to $1$). Considering the linear deterministic model of \cite{wnif} in Fig.~\ref{fig:fig2}-(b) for this network suggests that in a capacity achieving strategy each relay should carry information at rate approximately $\log a$ when $a$ is large. For partial decode-and-forward, the achievability strategy in the BC phase is superposition coding. (See Chapter~5 of \cite{abbasbook}.) The source generates three independent i.i.d. Gaussian codebooks of appropriate rates and powers and sends the addition of these three codewords. Each relay uses successive cancellation to decode its corresponding message: it successively decodes the codewords intended for the weaker relays and subtracts them from its signal in order to decode its own message while  codewords intended for the stronger relays are treated as additional noise. 

In our current example,  one natural choice for the powers of the superposed codebooks, to communicate  three messages of rates approximately $\log a$ to the three relays, can be $P_1=1/a^2$, $P_2=1/a$, and $P_3=1-1/a-1/a^2$. At large $a$, this corresponds to communication rates
\begin{align*}
R_1&=\log(1+a^3P_1)\approx \log a\\
R_2&=\log\left(1+\frac{a^2 P_2}{1+a^2P_1}\right)\approx \log a\,-1\\
R_3&=\log\left(1+\frac{a P_3}{1+a(P_2+P_1)}\right)\approx \log a\,-1
\end{align*}   
to the three relays. Note that there is a $1$ bit rate loss at each relay (except
for the strongest one) since the codebooks intended for the stronger relays
constitute additional noise at the weaker relays. In the corresponding
extension of this configuration to $N$-relays, this would result in $O(N)$ rate
loss between the sum broadcast rate to the relays and the capacity of the
single-input multiple output (SIMO) channel at the first stage, i.e. the cutset upper bound. (Note that the SIMO capacity is at least as large as the capacity of the strongest link, i.e. $\log(1+a^3)\approx 3\log a$ in our current example). 

The argument in the earlier section suggests that there should be a better way to choose
the broadcasting rates to the relays. For our current example, we can instead choose
$P_i=\frac{i}{a^{N-i}}$ for $i=1, \dots, N-1$ and $P_{N}=1-\sum_{i=1}^{N-1} P_i$ 
and obtain the rates
$$
R_1\approx \log a,\quad\dots,\quad R_{N-1}\approx \log a, \quad R_N\approx \log a-\log N.
$$
which also lie in the broadcast capacity region of the first stage. But in this
case, the sumrate is only $O(\log N)$ bits away from the SIMO capacity. This suggests that it is desirable to concentrate the hit due to superposition coding in the rate to the weakest relay. 

\section{Simulations}\label{sec:sim}
In the previous sections, we established an upper bound on the gap between the rate achieved by two strategies and the cutset-upper bound in the $N$-relay diamond
network.  These are worst case bounds over all possible channel configurations and SNR's. In this section, we aim to get a better understanding of the performance of these strategies and the tightness of the bounds via simulation results for different statistics of the channel  coefficients. We will focus on the partial decode-and-forward strategy (which was proven to have a worst case gap of $2\log N$ to the cutset upper bound) and compare it to simpler strategies such as using the best relay and amplify-and-forward. In the best relay strategy, only the relay with the largest end-to-end capacity is utilized; it decodes the message from the source and forwards it to the destination. In amplify-and-forward, each relay scales its received signal by an amount that satisfies the power constraint.  We examine two
variations of amplify-and-forward: when the relays forward an optimally scaled version of their received signal to maximize the end-to-end rate between $s$ and $d$ (each relay  does not necessarily transmit at full power), as well as having all relays
scale up their received signal to full power.   We call the second case naive amplify-and-forward. Amplify-and-forward is known to perform very well on the diamond network when all channel gains are equal to each other \cite{urssuhas}, so it is interesting to see how partial decode-and-forward compares to it under  common statistical models for the channel coefficients.

For our simulations, we consider a $10$-relay diamond network with a single
antenna at all nodes.  Since simulating the exact cutset upper bound in \eqref{eq:cutset} is difficult due to the optimization over the input distribution, we instead take \eqref{eq:cutset3} as the upper bound. Therefore our results provide an  upper bound on the actual gap. We simulate the channels for the low SNR regime ($\SNR = 1$) and the high SNR
regime ($\SNR = 1000$) under two different statistical models: Rayleigh and shadow fading. 

\begin{figure}[!t]
\centering
\includegraphics[width=7in]{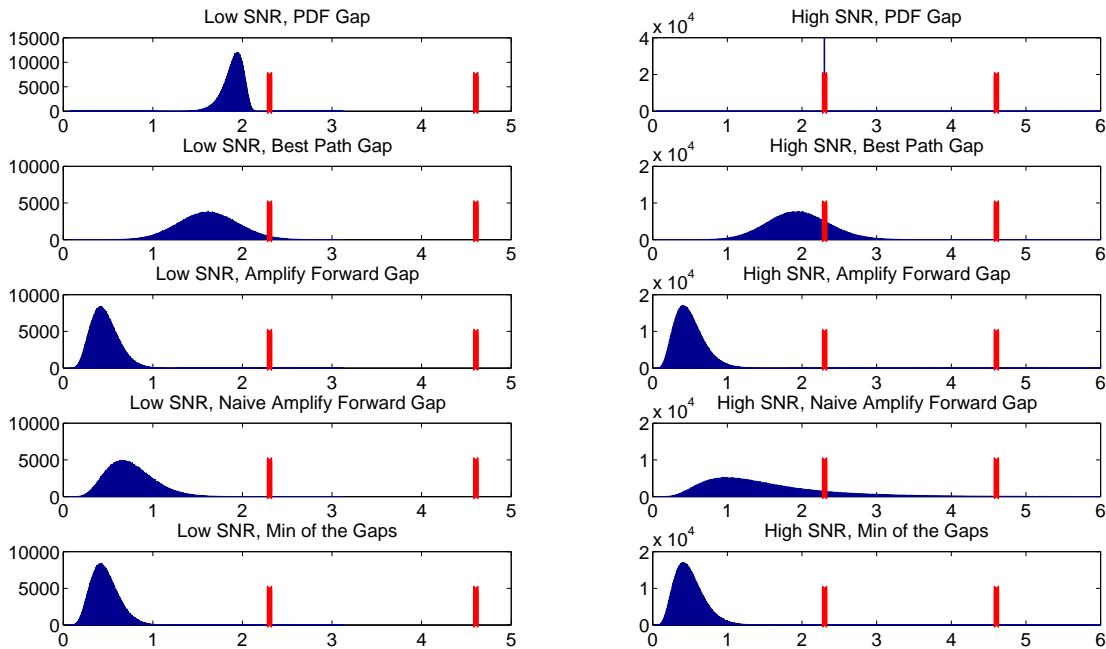}
\caption{The gap between various schemes with channel values being drawn $\mathbb{C}\N(0,1)$.  The two vertical lines represent $\log N$ and $2 \log N$.}
\label{fig:normal}
\end{figure}

\begin{figure}[!t]
\centering
\includegraphics[width=7in]{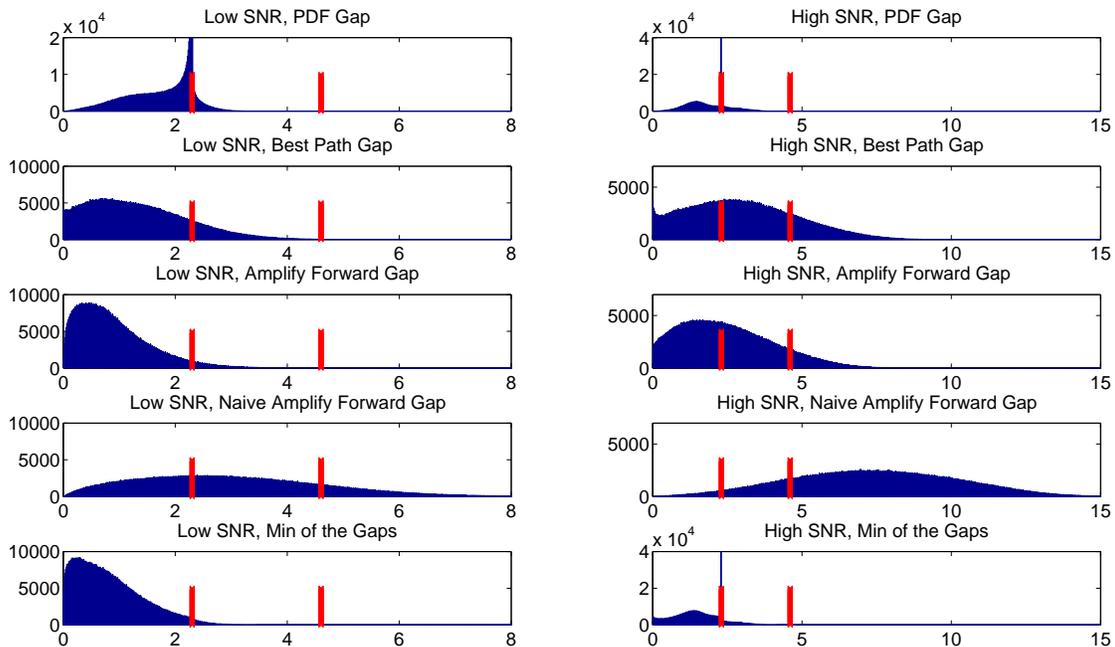}
\caption{The gap between various schemes with channel values being drawn assuming a shadowing model.  The two vertical lines represent $\log N$ and $2 \log N$.}
\label{fig:shadow}
\end{figure}

Figure \ref{fig:normal} shows a histogram of the gap between the cutset-upper
bound and the various schemes when the channel coefficients $h_{is}$ and
$h_{id}$ are drawn i.i.d.  $\mathbb{C}\N(0,1)$.  The last row is a histogram of
the gap when we take the best rate achieved among the four schemes for each
realization of the channel coefficients which represents an estimate of the remaining gap in the capacity of the $N$-relay diamond network (the difference  between the best achievability we have and the upper bound).  The two vertical lines in each plot
mark $\log N$ and $2 \log N$.  We note several features of the figure.  First,
the gap for partial decode-and-forward is always below $2 \log N$, as predicted
by our result.  In the high SNR regime, the gap is almost always at $\log N$.
On the other hand, amplify-and-forward has a much smaller gap both at high and low SNR. Even the simple scheme of only using the best relay seems to perform reasonably well. This is because with Rayleigh fading, there is limited variation between the channel gains. This favors amplify-and-forward, since it can obtain significant beamforming gain by coherently combining signals arriving over different paths. 

We also consider the case when we model the channel coefficients by shadowing,
where the channel attenuation in dB are drawn from a zero mean normal
distribution.  In other words, the channel coefficients, $h_{id}$ and $h_{is}$,
are distributed according to $10^{-\frac{X}{10}}$, and $X$ is a  normal
variable.  Typical standard deviations for this model range from $3$ - $14$
\cite{rappaport}.  In our simulations, we use a standard deviation of $7$.  A key
feature of shadowing is that some channel coefficients may be much larger than
others.  Figure \ref{fig:shadow} shows a histogram of the gap between the
cutset-upper bound and the various schemes under this model.  We
note some interesting differences between this new model and the previous model.

Partial decode-and-forward maintains a gap that is below $2 \log N$ independent
of the channel configuration and SNR (as predicted by our theoretical results).  On the other hand, while amplify-and-forward performed well under the earlier model, we now see that its gap can be quite large for some channel configurations. (Its performance becomes even worse if a we take a larger standard deviation for the shadowing model.) We also note that its performance is comparable to using only the best relay, as predicted in \cite{simplification}. Note that naive amplify-and-forward can have a very large gap, as relays  that are weak in the first stage can be injecting significant noise to communication when scaling up their received signals.  We conclude that while
amplify-and-forward can perform better than partial decode-and-forward in
certain channel configurations (most notably when channel gains are close to each other),  it cannot provide a universally good performance under all channel configurations.

\section{Conclusion}
In this paper, we developed a $O(\log N)$-bit approximation for the capacity of the $N$-relay diamond network, independent of the channel coefficients and the SNR, improving upon the existing $O(N)$-bit approximations for the capacity of this network.  
We showed that two strategies, noisy network coding and partial decode-and-forward can be optimized to achieve the information-theoretic cutset upper bound on the capacity of this network within $O(\log N)$ bits.  The discussion on noisy network coding reveals that the rule-of-thumb to quantize the received signals at the noise level used for compress-and-forward in the current literature can be highly suboptimal. Instead, it may be desirable for the relays to quantize at a much coarser scale. Extending our results to other topologies and deriving improved capacity approximations for general Gaussian relay networks remain as open problems with some initial results in this direction reported in \cite{ritesh}.

\begin{appendices}

\section{Diamond Network with Multiple Antennas}

\subsection{Proof Remark~\ref{rem:nnc}}
For the multiple antenna case, we choose $X_i, i \in \N$ to be i.i.d.
circularly symmetric Gaussian with covariance $\frac{P}{n_i}I$.  Let $X_s$ be
independent from $X_i$ and also circularly symmetric Gaussian with covariance
$\frac{P}{n_s}I$.  Also, we define $\hat{Y}$ to be such that
$$
\hat{Y}_i = Y_i + \hat{Z}_i, \qquad i \in \N,
$$
where $\hat{Z}_i$ are i.i.d. circularly symmetric Gaussian with covariance $\sigma^2(\sum_{i\in \N}n_i)I$ independent of everything else.  The first term in $R_\text{NNC}$ becomes
\begin{align*}
I(X_s, X_\L;\hat{Y}_{\ol{\L}}, Y_d | X_{\ol{\L}}) &= I(X_s; \hat{Y}_{\ol{\L}}) + I(X_\L; \sum_{i \in \L} H_{id}X_i + Z) \\
&= \log \det (I+\frac{\SNR }{n_s(\sum_{i \in \N} n_i +1)}\sum_{i\in\bar\L} H_{is}^\dagger H_{is}) + \log \det (I+\sum_{i\in\L} \frac{\SNR }{n_i} H_{id} H_{id}^\dagger).
\end{align*}
The second term is
\begin{align*}
I(Y_\L;\hat{Y}_\L|X_s,X_\L,\hat{Y}_{\ol{\L}},Y_d) &= \sum_{i=1}^{|\L|} n_i \log (1+\frac{1}{\sum_{i \in \N}n_i}) \\
&\leq \frac{\sum_{i=1}^{|\L|} n_i}{\sum_{i \in \N} n_i} \leq 1.
\end{align*}

from \eqref{eq:cutset3}, the cutset upper bound is bounded by 
\begin{align*}
\ol{C}&=\sup_{X_s,X_1,\dots, X_N} \min_{\L\subseteq \mathcal{N}} I(X_s, X_\L; Y, Y_{\ol{\L}}\,|\, X_{\ol{\L}}) \nonumber \\
&\leq \min_{\L \subseteq \N} \sup_{X_s} I(X_s; Y_{\ol{\L}}) + \sup_{X_\L} I(X_\L; \sum_{i \in \L} H_{id} X_i + Z) \nonumber \\
&\leq \min_{\L \subseteq \N} \Big( C_s(\ol{\L}) + C_d(\L) \Big),
\end{align*}
where $C_s(\ol{\L})$ is the MIMO capacity between the source node $s$ and
subset of relay nodes $\ol{\L}$ and $C_d(\L)$ is the MIMO capacity between
the remaining relay nodes $\L$ and the destination node $d$.  We can bound the
difference between the capacity of the MIMO channel under optimal power
allocation and under the equal power allocation on all antennas by applying the following
lemma:
\begin{lemma}\label{lemma1}
(Adapted from Appendix F in \cite{wnif}). Consider a MIMO channel with $n_t$
transmit antennas and $n_r$ receive antennas.  Let $C_\wf$ denote the capacity
of the channel under optimal power allocation, and let $C_\ep$ denote the
capacity of the channel under equal power allocation.  Then 
$$
C_\wf - C_\ep \le n\log \left(1 + \frac{n_t - 1}{n} \right),
$$
where $n = \min(n_r, n_t)$.
\end{lemma}
The proof of the lemma is given at the end of the appendix.

Since the source has power $P$, equal power allocation among the $n_s$ transmit antennas of the source yields
$$
C_{s,\ep}(\ol{\L}) = \log \det \Big(I + \frac{\SNR}{n_s}\sum_{i\in \ol{\L}} H_{is}^\dagger H_{is} \Big),
$$
for the MIMO channel MIMO between $s$ and
 $\ol{\L}$ and so
\begin{align*}
C_s(\ol{\L}) &\leq \log \det \Big(I + \frac{\SNR}{n_s}\sum_{i\in \ol{\L}} H_{is}^\dagger H_{is} \Big) + \min(n_s, \sum_{i \in \ol{\L}} n_i) \log (1 + \frac{n_s -1}{\min(n_s, \sum_{i \in \ol{\L}} n_i)}) \\
&\leq \log \det \Big(I + \frac{\SNR}{n_s}\sum_{i\in \ol{\L}} H_{is}^\dagger H_{is} \Big) + n_s \log (1 + \frac{n_s -1}{n_a}),
\end{align*}
where we define $n_a = \min(n_s, n_1, \ldots, n_N)$ and use the fact that $n_s\leq \sum_{i\in\mathcal{N}}n_i$.  Similarly, for $C_d(\L)$, we have
total power $|\L|P$ among $\sum_{i \in \L} n_i$ transmit antennas, so
\begin{align*}
C_d(\L) &\leq \log \det \Big(I + \frac{|\L|\SNR}{\sum_{i \in \L} n_i} \sum_{i\in \ol{\L}} H_{id} H_{id}^\dagger \Big) + \min(n_d, \sum_{i \in \L} n_i) \log (1 + \frac{\sum_{i \in \L} n_i -1}{\min(n_d, \sum_{i \in \L} n_i)}) \\
&\leq \log \det \Big(I + \frac{\SNR}{\min_{i \in \L} n_i}\sum_{i\in \L} H_{id} H_{id}^\dagger \Big) + n_d \log (1 + \frac{\sum_{i \in \L} n_i - 1}{n_b}),
\end{align*}
where we define $n_b = \min(n_d, n_1, \ldots, n_N)$ and use the fact that $n_d\leq \sum_{i\in\mathcal{N}}n_i$.  Thus, we can upperbound the cutset bound as
\begin{align}
\ol{C}\leq \min_{\L \subseteq \N} \Bigg(\log \det \Big(I + \SNR \sum_{i\in \ol{\L}} H_{is}^\dagger H_{is} \Big) &+ \log \det \Big(I + \SNR \sum_{i\in \L} H_{id} H_{id}^\dagger \Big) \Bigg)\\
  &+ n_s \log (1 + \frac{n_s -1}{n_a}) + n_d \log (1 + \frac{\sum_{i \in \N}n_i - 1}{n_b}).\label{eq:ubcutset}
\end{align}
The gap between the cutset-upper bound and $R_{NNC}$ is then upper bounded by
\begin{equation}
R_{NNC} \ge \ol{C} - G_1,
\end{equation}
where $G_1 = n_s \log \left( \sum_{i \in \N} n_i + 1\right) + n_s \log (1 +
\frac{n_s -1}{n_a}) + n_d \log (\max_{i \in \N} n_i) + n_d \log (1 +
\frac{\sum_{i \in \N}n_i - 1}{n_b})$, and $n_a = \min(n_s, n_1, \ldots, n_N)$,
$n_b = \min(n_d, n_1, \ldots n_N)$.

\subsection{Proof Remark~\ref{rem:pdf}}

To determine the rate achieved by partial decode-and-forward with multiple antennas, we identify the
set of rates $(R_1, \ldots, R_N)$ that lie in the intersection of the BC and
MAC capacity regions and find the largest sum rate $\sum_{i=1}^N R_i$.  As with
the scalar case, we lower bound this rate by finding polymatroidal subregions
of the BC and MAC capacity regions and applying Edmond's polymatroidal
intersection theorem.

The capacity region for the MIMO MAC with user $i$ having average power
constraint
$P_i$ is given by \cite{wireless}: $\C_\MAC(P_1, \ldots, P_N, H_{1d}, \ldots, H_{Nd})=$
\begin{align*}
 \bigcup_{\{\tr(Q_i) \leq P_i \sp \forall i \}} \Bigg\{ (R_1, \ldots R_N) : \sum_{i \in S} R_i \leq \log \det \left(I + \sum_{i\in S} H_{id} Q_i H_{id}^\dagger \right), \forall S \subseteq \N, R_i \ge 0 \ \forall i \Bigg\},
\end{align*}
where $H_{id}$ is the $n_d \times n_i$ channel matrix between user $i$ and the
destination.  The duality between the capacity regions of the MIMO BC and MIMO
MAC \cite{bcmac} yields a characterization of the MIMO BC region in terms of
the MIMO MAC capacity:
$$
\C_\BC(P, H_{1s}, \ldots, H_{Ns}) = \bigcup_{\sum_i P_i \le P} \C_\MAC(P_1, \ldots, P_N, H_{1s}^\dagger, \ldots, H_{Ns}^\dagger),
$$
where $H_{is}$ is the $n_i \times n_s$ channel matrix between the source $s$ and
receiver $i$ in the BC channel.  We now identify polymatroidal subregions of the MIMO MAC and MIMO BC capacity regions.

For the diamond relay network, each relay has power constraint $P$, so for the
MIMO MAC region, we choose $Q_i = \frac{P}{n_i}I$ to have equal power among the
$n_i$ antennas for each relay, thus yielding a subregion of the MIMO MAC
capacity, $P(f) \subseteq \C_\MAC$, where
$$
P(f)=\{(R_1,\dots, R_N): \sum_{i\in S} R_i\leq f(S), \forall S\subseteq\mathcal{N}, R_i\geq 0,\, \forall i\}, \\
$$
and
$$
f(S) = \log \det \left(I + \sum_{i\in S} \frac{\SNR}{n_i}H_{id}H_{id}^\dagger \right).
$$
The function $f$ satisfies the conditions in Definition \ref{def1}, and so
$P(f)$ is a polymatroid.

For the MIMO BC capacity, we apply the BC MAC duality with $P_i = n_iP/\sum n_i$ and
$Q_i = \frac{P}{\sum n_i}I$, which gives $P(g) = \C_\MAC(n_1P/\sum n_i,
\ldots, n_NP/\sum n_i, H_{1s}^\dagger, \ldots, H_{Ns}^\dagger) \subseteq
\C_\BC$, where
$$
P(g)=\{(R_1,\dots, R_N): \sum_{i\in S} R_i\leq g(S), \forall S\subseteq\mathcal{N}, R_i\geq 0,\, \forall i\}, \\
$$
and
$$
g(S) = \log \det \left(I + \frac{\SNR}{\sum_{i \in \N} n_i} \sum_{i\in S} H_{is}^\dagger H_{is} \right).
$$
$P(g)$ is also a polymatroid, so as with the single antenna case, we apply
Edmond's polymatroid intersection theorem to get
\begin{align*}
R_{PDF}&= \min_{\L \in \N}\quad f(\L) + g(\ol{\L}) \\
&= \min_{\L \in \N} \Bigg(\log \det \left(I + \frac{\SNR}{\sum_{i \in \N} n_i} \sum_{i\in \ol{\L}} H_{is}^\dagger H_{is} \right) + \log \det \left(I + \sum_{i\in \L} \frac{\SNR}{n_i}H_{id} H_{id}^\dagger \right) \Bigg) \\
&\ge \min_{\L \in \N} \Bigg(\log \det \left(I + \frac{\SNR}{\sum_{i \in \N} n_i} \sum_{i\in \ol{\L}} H_{is}^\dagger H_{is} \right) + \log \det \left(I + \frac{\SNR}{\max_{i \in \N} n_i} \sum_{i\in \L} H_{id} H_{id}^\dagger \right) \Bigg).
\end{align*}
By comparing it to \eqref{eq:ubcutset}, it can be verified that 
\begin{equation*}
R_{PDF} \ge \ol{C} - G_2.
\end{equation*}
This completes the proof of Remark~\ref{rem:pdf}.

\subsection{Proof of Lemma~\ref{lemma1}}
Suppose we have a MIMO channel with $n_t$ transmit antennas, $n_r$ receive
antennas, and total power $n_tP$.  Let $n = \min(n_r, n_t)$.  The capacity of
the MIMO channel is well known to be
$$
\C_\wf = \sum_{i=1}^n \log (1 + \tilde{Q}_{ii}\lambda_i),
$$
where the $\lambda_i$ correspond to the singular values of the MIMO channel
matrix and $\tilde{Q}_{ii}$ is given by the waterfilling solution satisfying
$$
\sum_{i=1}^n \tilde{Q}_{ii} = n_tP.
$$
The rate achieved by equal power allocation is
$$
\C_\ep = \sum_{i=1}^n \log (1 + P\lambda_i).
$$
Assume without loss of generality $\lambda_1 \ge \lambda_2 \ge \ldots \ge
\lambda_n$.  We upperbound $\C_\wf - \C_\ep$ as follows:
\begin{align*}
\C_\wf - \C_\ep &= \log \left(\frac{\prod_{i=1}^n (1 + \tilde{Q}_{ii}\lambda_i)}{\prod_{i=1}^n (1 + P\lambda_i)} \right) \\
&= \log \prod_{i=1}^n \left(\frac{1 + \tilde{Q}_{ii}\lambda_i}{1+P\lambda_i} \right) \\
&\stackrel{(a)}{\leq} \log \left( \frac{1}{n}\sum_{i=1}^n \Big(\frac{1 + \tilde{Q}_{ii}\lambda_i}{1+P\lambda_i} \Big) \right)^n\\
&= n\log \left( \frac{1}{n} \Big(\sum_{i=1}^n\frac{1}{1+P\lambda_i}+ \sum_{i=1}^n\frac{ \tilde{Q}_{ii}\lambda_i}{1+P\lambda_i} \Big) \right) \\
&= n\log \left( \frac{1}{n} \Big(\sum_{i=1}^n\frac{1}{1+P\lambda_i}+ \sum_{i=1}^n\tilde{Q}_{ii}(\frac{1}{P} - \frac{1}{P(1+P\lambda_i)}) \Big) \right) \\
&\stackrel{(b)}{\leq}  n\log \left( \frac{1}{n} \Big(\sum_{i=1}^n\frac{1}{1+P\lambda_i}+ \sum_{i=1}^n\tilde{Q}_{ii}(\frac{1}{P} - \frac{1}{P(1+P\lambda_1)}) \Big) \right) \\
&=  n\log \left( \frac{1}{n} \Big(\sum_{i=1}^n\frac{1}{1+P\lambda_i}+ \frac{n_t P \lambda_1}{1+ P\lambda_1 } \Big) \right) \\
&=  n\log \left( \frac{1}{n} \Big(\sum_{i=2}^{n}\frac{1}{1+P\lambda_i}+ \frac{1+ n_t P \lambda_1}{1+ P\lambda_1 } \Big) \right)
\end{align*}
\begin{align*}
&\stackrel{(c)}{\leq}  n\log \left(\frac{n-1}{n}+ \frac{1+ n_t P \lambda_1}{n(1+ P\lambda_1 )} \right) \\
& = n\log \left(\frac{n-1}{n}+ \frac{1}{n}\Big(n_t - \frac{n_t -1}{1+ P\lambda_1}\Big) \right) \\
&\stackrel{(d)}{\leq}  n\log \left(1 + \frac{n_t-1}{n} \right). 
\end{align*} 
where (a) follows from the arithmetic mean-geometric mean inequality, (b) follows from the fact that $\lambda_1 \ge \lambda_2 \ge \ldots \ge
\lambda_n$, (c) follows from $\frac{1}{1+P\lambda_i}\leq 1$ and (d) is obtained by discarding the last term in the previous line.
\end{appendices}
\end{document}